# CALIPSO observations of wave-induced PSCs with near-unity optical depth over Antarctica in 2006-2007

Noel V.[1], Hertzog A.[2] and Chepfer H.[2]

[1] *Laboratoire de Météorologie Dynamique, Institut Pierre-Simon Laplace, CNRS, Palaiseau, France*

[2] *Laboratoire de Météorologie Dynamique, UPMC Univ. Paris 06, Palaiseau, France*

Contact:	vincent.noel@lmd.polytechnique.fr

Vincent Noel

Laboratoire de Météorologie Dynamique

Ecole Polytechnique

91128 Palaiseau, France

Tel: 33169335146

**Abstract**

Ground-based and satellite observations have hinted at the existence of Polar Stratospheric Clouds (PSCs) with relatively high optical depths, even if optical depth values are hard to come by. This study documents a Type II PSC observed from spaceborne lidar, with visible optical depths up to 0.8. Comparisons with multiple temperature fields, including reanalyses and results from mesoscale simulations, suggest that intense small-scale temperature fluctuations due to gravity waves play an important role in its formation; while nearby observations show the presence of a potentially related Type Ia PSC further downstream inside the polar vortex. Following this first case, the geographic distribution and microphysical properties of PSCs with optical depths above 0.3 are explored over Antarctica during the 2006 and 2007 austral winters. These clouds are rare (less than 1% of profiles) and concentrated over areas where strong winds hit steep ground slopes in the Western hemisphere, especially over the Peninsula. Such PSCs are colder than the general PSC population, and their detection is correlated with daily temperature minimas across Antarctica. Lidar and depolarization ratios within these clouds suggest they are most likely ice-based (Type II). Similarities between the case study and other PSCs suggest they might share the same formation mechanisms.

## 1. Introduction

Extinction at visible wavelengths is generally considered very small in Polar Stratospheric Clouds (PSCs). Using limb measurements from the SAM II spaceborne sun photometer, *McCormick and Trepte* (1987) established a range of 0.001-0.06 for stratospheric optical depth at 1.0 μm, peaking at 0.026 above Antarctica, but these weekly averages are difficult to relate to local optical depths of individual PSCs. Since then, PSC climatologies based on limb and nadir satellite observations have provided limited information about optical depth; moreover, 1) radiances from passive instruments in polar orbit have trouble detecting optically thin clouds (*Hervig et al.* 2001), and 2) large differences exist between results from instruments (*Pavolonis and Key* 2003). *Wang and Michelangeli* (2006) used extinction profiles retrieved over the Arctic from the Improved Limb Atmospheric Spectrometer; even assuming 10-km thick PSC, these profiles translate to optical depths smaller than 0.01. *Höpfner et al.* (2001), through the analysis of the solar absorption spectrum, found optical depths 0.25-0.8 in the infrared (10.6-12.5 μm) for a case study PSC over Sweden, a result difficult to extend to the visible domain. Regarding lidar observations, *Reichardt et al.* (2004) derived optical depths as part of their analysis of mountain wave PSCs over Sweden, but were not presented in the text. Due to this lack of quantitative reference, the 2002 study of Arctic stratospheric components by *Tabazadeh et al.* still refers to an optical depth range below 0.04 for PSCs, as documented in *McCormick et al.* 1981.

Despite the scarcity of actual values in literature, PSCs with relatively high optical depths are not unheard of – for instance over the Arctic using the Michelson Interferometer for Passive Atmospheric Sounding (MIPAS) in *Spang et al.* (2005), or from abnormally high profile terminations in long-term datasets from photometer and occultation instruments (*Fromm et al.* 2003*, Alfred et al.* 2007) – but these reports are mostly anecdotal. Higher optical depths in PSCs could be due to either bigger or more numerous particles, in which case such clouds would lead to enhanced stratospheric denitrification, slowing down polar ozone recovery (*Carslaw et al.* 1998*, Jensen et al* 2002). As global model simulations indicate future climate change might increase PSC optical depth by a factor of 5 (*Pitari et al.* 2002) and lead to increase in polar greenhouse effect, a representative assessment of PSC optical depth is important to climate prediction. In this regard, the present study aims at evaluating the importance and occurrence of PSCs with relatively high optical depths (above 0.3) over Antarctica.

Results from the Geosciences Laser Altimeter System (*Spinhirne et al.* 2005) show the sensitivity of spaceborne lidars make them very well suited to the reliable detection of PSCs (*Palm et al.* 2005a, 2005b). Here, we present observations of PSCs with optical depths above 0.3 from the spaceborne lidar CALIOP (Cloud-Aerosol Lidar with Orthogonal Polarization) during the 2006 and 2007 polar winters above Antarctica. CALIOP is part of the CALIPSO mission (Cloud Aerosol Lidar and Infrared Pathfinder Satellite Observations), and orbits Earth between 82°N and 82°S at 705 km (*Winker et al.* 2007). Following previous PSC studies using CALIOP (*Pitts et al.* 2007; *Noel et al.* 2008), only nighttime observations were used in the present paper, as they describe most of the Antarctic wintertime and offer higher signal-to-noise ratio than daytime observations.

After introducing the dataset and the algorithms used for cloud detection (Sect. 2), a case of PSC with near-unity optical depth is described and analyzed against multiple temperature fields (Sect. 3), shown to be generated by gravity-wave temperature fluctuations. Occurrence and properties of PSCs over 2006 and 2007 winters are described in Sect. 4. Results are summarized and their implications for PSC formation discussed in Sect. 5.

## 2. CALIOP Data and Cloud Detection

The present study uses observations of perpendicular and attenuated total backscatter at 532 nm as a function of altitude, longitude and latitude. The depolarization ratio $\delta$ (*Sassen* 1991) was computed as the ratio between perpendicular and parallel (total minus perpendicular) backscatter. The lidar ratio S was computed, as in *Noel et al.* (2007), by the ratio between anisotropic extinction and backscattering coefficients, both integrated over the cloud layer.

CALIOP observations are provided on a non-uniform altitude and time grid (*Winker et al.* 2007), with a vertical resolution between 30 m and 300 m. We re-gridded the data over uniform altitude bins of 30 meters and performed an average of successive 30 profiles, reducing the horizontal resolution to 10 km in order to increase signal-to-

4                                                                                                   4

noise ratio. We then normalized the 532 nm attenuated total backscatter on the molecular backscattering profile in the middle stratosphere (between 32 and 34 km, above possible PSC layers, similar to *Innis and Klekociuk* 2006). To detect clouds, we imposed a minimum threshold on attenuated total backscatter above molecular backscattering, as sensitivity studies showed this criteria gives the best results compared to imposing a minimum threshold on backscatter, scattering ratio or depolarization ratio. A threshold of $8 \times 10^{-4}$ km$^{-1}$sr$^{-1}$ gives results consistent with ground-based observations (*Immler et al.* 2007) and independent CALIOP validations (*McGill et al.* 2007); extensive comparisons show that these technique and value provide a successful detection of PSCs at least on par with *Noel et al. 2008*. In order to reduce the number of false positives in cloud detection due to instrumental noise, profiles with less than 5 consecutive cloud points were flagged as clear sky, profiles with 5 to 20 consecutive cloud points as unidentified, and profiles with more than 20 consecutive cloud points as cloudy (i.e. clouds at least 600 m tall). Only these last profiles were considered in the following results. Moreover, "transition" profiles (i.e. a profile between a cloudy and a clear-sky profile) were also excluded. For a given atmospheric profile, only cloud points above the tropopause were considered as PSCs; tropopause heights were extracted from the CALIOP level 2 data products (v.1.20 if available, 1.10 otherwise), provided by NASA's Global Modeling and Assimilation Office (GMAO) using the Goddard Earth Observing System Model 5 (GEOS-5 model). Applying this detection scheme on the CALIOP dataset correctly tracks the wintertime increase and decrease in stratospheric cloud cover over Antarctica, from near-zero mid-May to its July peak and back near zero in late September; this gives confidence the number of false detections is low.

Within cloudy lidar profiles, we looked for a 1-km clear-sky area (33 points) from the tropopause and up. If clear sky was not found below 28 km the profile was rejected; otherwise PSC extinction and optical depth τ were estimated by the difference in molecular and attenuated backscatter within the clear sky area (*Young* 1995, also used in *Reichardt et al.* 2004). Multiple scattering effects should be negligible considering the low extinctions under consideration (*Chepfer et al.* 1999); these effects could be significant for highly opaque PSCs, but would lead to underestimated optical depths and would not change the conclusions.

### 3. Case study : June 27th 2006

3.1 Lidar observations and optical depth

PSCs are traditionally categorized as Type Ia (HNO$_3$-based particles like nitric acid trihydrate or NAT), Type Ib (supercooled ternary H$_2$SO$_4$/HNO$_3$/H$_2$O solution or STS) or Type II (ice-based). Lidar observations have been used qualitatively for several years now to discriminate them, for instance Type II PSCs generally produce stronger lidar backscatter (*Hopfner et al.* 2006a). Several schemes have been devised for automatic classification, from the distinction of the three PSC types using depolarization and scattering ratios at one wavelength (*Browell et al.* 1990) or several (*Dornbrack et al.* 2002, *Hu et al.* 2002), to attempts to identify sub-types of HNO$_3$-based particles like NAT-rock, enhanced NAT or NAD (e.g. *Adriani et al.* 2004, *Massoli et al.* 2006). Here we use lidar observations to provide a qualitative assessment of microphysical properties in the observed PSCs.

Fig. 1 shows CALIOP observations as a function of latitude and altitude for the June 27th 2006 orbit (03:28UTC segment), with CALIPSO's trajectory plotted over Antarctica as an inset. Several layers of PSCs were identified between 18 to 28 km over the polar region in this orbit; for presentation purposes only the segment between 60 and 75°S is shown. Attenuated total backscatter (top panel, Fig. 1a) shows a homogeneous and bright PSC layer from 69°S to 74°S (red orbit section in the inset). This PSC extends over at least 600 km along CALIPSO's orbit, eastward of the Antarctic Peninsula (i.e. downstream according to the direction of the polar vortex wind). The integrated total backscatter between 14 and 28 km is close to $10^{-2}$ km$^{-1}$.sr$^{-1}$. Depolarization ratio after clear-sky removal is shown in Fig. 1b - dimmer PSCs (68 to 63°S, 20 to 25 km, light blue shading in Fig. 1a) are correctly not picked up. Lidar ratios S are shown in Fig. 1c.

Following *Adriani et al.* (2004), $\delta > 0.3$ should be indicative of solid cristalline composition; *Reichardt et al.* (2004) report $56 < S < 135$ for Type I and $16 < S < 42$ for ice-based Type II PSCs (*Platt et al.* (1999) report a 25-34 range for cirrus clouds). In the present case, high depolarization ratios ($0.3 < \delta < 0.55$) and low lidar ratios (S ~ 20, Fig. 1c) are consistent with Type II. Optical depths τ (Fig. 1d) are significantly greater than 0.5; by comparison, τ < 0.2 for the dimmer PSC between 66°S and 69°S.

## 3. 2. Comparison of CALIOP observations with temperature fields

Nucleation and particle growth occurring in PSCs are still not precisely understood (*Brooks et al.* 2004), especially in Type Ia (*Wang and Michelangeli* 2006), where several formation processes compete (*Carslaw et al.* 1999). However, it is clear that PSCs form in cold temperatures, either during synoptical scale cooling or local fluctuations due to gravity waves (*Teitelbaum et al.* 2001). Temperature thresholds for particle nucleation depend on stratospheric concentrations of aerosols and water vapour; under normal ambient stratospheric conditions at 20 km, NAT particles (Type Ia) appear to condense near 195 K, STS particles (Ib) near 191 K and ice particles (II) at or below 188 K (*Alfred et al.* 2007). To evaluate the validity of the microphysical interpretation of the present case (Sect. 3.1), temperature fields were extracted for the area and time under study from large-scale global models and reanalyses:

- Temperatures provided in the CALIOP level 1 data files (GMAO), in two versions: GEOS-4 (CALIOP level 1 version 1), and GEOS-5 (CALIOP level 1 version 2) that is more accurate in polar regions (*Thomason et al.* 2007).

- 00UTC temperatures from the National Centers for Environmental Prediction (NCEP) at 2.5° resolution and 17 pressure levels (1000-10 hPa) (*Kistler et al.* 2001).

- 00UTC temperature from the European Center for Medium-Range Forecast (ECMWF) operational analyses (*Rabier et al.* 2000) retrieved at 0.5° resolution on the 21 pressure-level archive (1000-1 hPa).

Moreover, the Weather and Research and Forecasting model (WRF v.2.2, *Skamarok et al.* 2007) was used to simulate thermodynamical conditions around the observed PSC. The simulated domain was a 100x100 grid at 20 km resolution centered on the Antarctic Peninsula with 120 vertical levels. A 5 km thick damping layer was set right below the top of the simulated atmosphere, in order to avoid reflection on the model ceiling (e.g. *Watanabe et al.* 2006). Three simulations used different minimum pressure levels (defining maximum altitude) and meteorological data for initialization and boundary conditions (Table 1).

|      | Meteorological input data and resolution | Minimum pressure level, maximum altitude |
|------|------------------------------------------|------------------------------------------|
| WRF1 | NCEP, 2.5°                               | 10 hPa (~28 km)                          |
| WRF2 | ECMWF, 0.5°                              | 10 hPa (~28 km)                          |
| WRF3 | ECMWF, 0.5°                              | 4 hPa (34 km)                            |

Table 1 - Meteorological data and minimum pressure levels for WRF simulations

NCEP data were not available above the 10 hPa level. Simulations were initiated on June 25th, 2006; results were extracted at 03UTC on June 27th 2006 for consistency with CALIOP and GMAO results.

Temperatures from these sources were projected on CALIOP coordinates and altitude (Fig. 2). For visual comparison, a contouring filter was applied to CALIOP data, outlining areas with backscatter higher than $1.5 \cdot 10^{-3}$ km$^{-1}$.sr$^{-1}$. GMAO data are natively interpolated on the CALIOP grid, which explains their smoothness. Temperatures from GMAO-GEOS4 (Fig. 2a) and NCEP (Fig. 2c) are homogeneous and free of small-scale fluctuations (as in *Gobiet et al.* 2005). Temperature stratification is disturbed in results from GMAO-GEOS5 (Fig. 2b), but without significant change in actual temperatures. ECMWF shows spatial inhomogeneities (Fig. 2d), and large fluctuations above the Antarctic peninsula (~70°S for this particular CALIPSO orbit track), down to -100°C. Small-scale variations appear on results from WRF1 (Fig. 2e) and WRF2 (Fig. 2f), but cold temperatures are not correlated well with the cloud observed by CALIOP, possibly because in those configurations the cloud top is located within the damping layer. Results from WRF3 (Fig. 2g), where the damping layer begins above the cloud top, show an extremely good geographic and altitude correlation with the PSC detected by CALIOP. This correlation, and the fact that this last simulation is the less prone to numerical side-effects due to its high damping layer, suggests the WRF3 results are the most physically realistic. Besides, comparing this figure with Fig. 1 shows that the optically thin PSCs present between 20 and 24 km between 68°S and 63°S (not targeted by the present study) also seem to follow the cold temperature patterns evidenced by WRF3, as do other PSCs from the same orbit absent from the figures (i.e. poleward of 75°S).

Temperature distributions within the observed PSC from the same sources are shown in Fig. 3. Temperatures using WRF3 (Fig. 3e) are the coldest (-100°C to -85°C, 173 to 188K), and the distribution is the narrowest, showing the reached accuracy is able to create a homogeneous field within a very precisely constrained area. These temperatures are colder than the ice frost point (roughly 188 K, *Voigt et al.* 2000), and are consistent with an identification as a Type II PSC (Sect. 3.1). Results from ECMWF, WRF1 and WRF2 (Fig. 3b to 3d) include similar cold temperatures, but probably due to misalignment, they also include warmer temperatures (up to -70°C, 203K), increasing the dispersion and resulting in large distributions. Distributions from NCEP and GMAO (Fig. 3a and 3b) are narrow, but this is because these models do not reproduce well the wave-induced temperature disturbances; temperatures are overall warmer, -90°C to -80°C (183 to 173K). GEOS-5 results are actually warmer than from GEOS-4 (Fig. 3a). When looking at distribution modes, results from GMAO are 10K warmer than those from WRF3.

### 3.3 Dynamical interpretation

Figure 4 displays horizontal maps and zonal cross-sections of the temperature and zonal-velocity fields simulated in WRF3 experiment on June 27, 0300 UT. The last column of that figure furthermore shows the disturbances of temperature and zonal velocity, which have been obtained by removing a second-order polynomial fit to the simulated fields at each latitude in the WRF domain. This figure highlights the presence of large-amplitude disturbances located above and directly in the lee of the Antarctic Peninsula. The cross sections furthermore reveal that these disturbances are observed throughout the atmosphere from the ground to the stratosphere. This kind of meso-scale feature is typical of a gravity-wave packet generated by the tropospheric flow passing over the Peninsula and propagating upward in the atmosphere. One can also notice that the shape of the PSC in Figure 3 is explained by the westward tilt with altitude of the mountain-wave phase fronts in the stratosphere. The Antarctic Peninsula, which is essentially a mountain ridge perpendicular to the prevailing westerly winds blowing over the Southern Ocean, has already been reported as a favored location for the generation of orographic waves in a number of studies (e.g., *Gary 1989*, *Alexander and Teitelbaum 2007*, *Plougonven et al. 2008*). Finally, a WRF simulation performed by setting the ground elevation to zero everywhere in the domain makes the temperature variations disappear in the stratosphere, confirming the orographic origin of these disturbances.

The amplitudes of the mountain wave simulated in WRF3 are particularly impressive. For instance, at 25 km (about 20 hPa) and 72°S, the peak-to-peak amplitudes of temperature and zonal-velocity disturbances locally reach 30 K and 120 m s$^{-1}$, respectively. On the other hand, the disturbances in meridional velocity are typically less than 10 m s$^{-1}$, indicating that the mountain wave is primarily zonally propagating. ECMWF analyses captured fairly well the amplitude of the mountain wave, most likely due to its relatively large scales. For comparison, *Gary* (1989) reported peak-to-peak amplitudes in temperature disturbances of the order of 10 K in the western side of the Peninsula, while *Plougonven et al.* (2008), who studied a mountain-wave event in the same location than in the present study, reported peak-to-peak amplitudes of 20 K and 20 m s$^{-1}$ in temperature and zonal velocity, respectively.

In the stratosphere, the horizontal ($\lambda_h$) and vertical ($\lambda_z$) wavelengths of the wave respectively amount to 400 km and 13 km (evaluated by visually inspecting slices of WRF3 thermodynamical variables at several latitudes and altitudes). As shown in the horizontal maps, the wave fronts are parallel to the mountain ridge (i.e. meridionally-aligned), so that, as already mentioned, the wave horizontal phase speed is essentially zonal. Using the gravity-wave dispersion relation and an estimation of the wavelengths of the wave packet, one can infer the wave intrinsic period ($\hat{T}$), i.e. the period of the fluctuations in the frame of reference moving with the flow:

$$\hat{T} = \frac{2\pi}{N} \frac{\lambda_h}{\lambda_z} \approx 2 \text{ h } 45 \text{ min} \qquad (1)$$

where $N$, the buoyancy frequency, is typically 2 10$^{-2}$ rad s$^{-1}$ in the lower stratosphere. This period is the one felt by air parcels passing over the mountain, while on the other hand mountain waves are typically stationary in the frame of reference linked to the ground.

The maximum cooling/heating rates associated with this wave packet can be computed from the intrinsic period and the amplitude of the temperature disturbances:

$$\frac{DT}{Dt} = \frac{2\pi T'}{\hat{T}} \approx \pm 34 \text{ K h}^{-1} \qquad (2)$$

where $T$ is the temperature, and $T'$ the amplitude of the temperature disturbance produced by the wave. This value is typical of mesoscale disturbances (e.g., *Hertzog et al., 2002*), and exceeds by one order of magnitude the heating/cooling rates implied by planetary-scale features in the lower stratosphere.

### 3.4. Observation of a downstream Type I PSC

Inspection of nearby CALIPSO orbits reveals, downstream with respect to polar vortex winds (eastward, previous descending orbit, Fig. 5 bottom panel), a dimmer, low-backscatter, weakly depolarizing ($0.05 < \delta < 0.2$) PSC. Such optical properties suggest a NAT-based composition (Type Ia). By contrast, observations from the westward, consecutive descending CALIPSO orbit show clear sky upstream (Fig. 5, middle panel). Isentropic back-trajectories from four points inside the downstream PSC (numbers in Fig. 5), computed using ECMWF wind fields, show these air masses were subject to temperature fluctuations associated with mountain waves in the vicinity of the Peninsula (Fig. 5, top panel), either on the location corresponding to the ice PSC on Fig. 1 (i.e., near 70-75°S and 60°W), or to the south (near 77°S and 70°W). In each case, back-trajectories crossed areas where wave-induced fluctuations led to temperatures colder than the frost point $T_{ICE}$ (obtained in Fig. 5 by assuming a water vapor mixing ratio of 5 ppmv in the polar stratosphere, using the *Marti and Mauersberger* (1993) formula). The geographical extension of temperatures below $T_{ICE}$ encountered by back-trajectory #3 is nevertheless very small (and hardly visible on Fig. 5). However, the simulated absolute temperatures in WRF3 may be slightly biased: temperatures along the ECMWF back-trajectory exhibit a drop to 178 K where back-trajectory #3 encountered temperature just below $T_{ICE}$ in the WRF3 simulations.

CALIOP backscatter observations made two days later roughly over the same domain (Fig. 6) still show a Type II PSC in the same location and altitude as the case study, with very similar shape and optical properties (orbit 03-15-58, Fig. 6 middle panel).

Since according to the WRF simulations the gravity wave was still in effect on June 29th (Fig. 6, top and right panels), resulting in two days of temperatures compatible with Type II PSC persistence, it can be assumed, barring any exceptional coincidence, that these are sightings of the same PSC lasting more than 48h. Moreover, observations from the previous CALIPSO descending orbit still detect a Type I PSC downstream with respect to polar vortex winds (orbit 01-37-03, Fig. 6 bottom panel), while upstream is still clear-sky (not shown).

**4. PSCs with τ > 0.3 in the 2006 and 2007 Antarctic winters**

Following the retrieval of near-unity optical depths for the case study, the techniques used to identify clouds and retrieve their optical depth (Sect. 2) were applied on CALIOP profiles between 60°S and 82°S over two consecutive austral winters, beginning June 16th (when CALIOP data acquisition started) in 2006 and May 1st in 2007 until the end of September (*Winker et al.* 2007). PSCs with τ > 0.3 were identified, as an optical depth of 0.3 was used as the lower boundary for opaque cirrus clouds by *Sassen and Cho* (1992).

4. 1. Frequencies and location

PSCs with τ > 0.3 appear in less than 1% of profiles, primarily in the western hemisphere above the continent (70 to 80°S, 30 to 120°W), often in a narrow stretch directly above the Antarctic Peninsula (Fig. 7). In this last area, such PSCs were detected as far North as 60°S. These findings are very consistent in 2006 and 2007. These PSCs appear lined up, thus in individual orbits, which suggests a limited lifetime or a fast decrease in optical depth between consecutive orbits. An isolated section of PSCs appears between 170°W and 180°W in 2007, in the lee of the Transantarctic mountain; inspection of data shows it is a genuine PSC spanning several orbits occurring on August 3rd, 2007. Time-wise, PSCs with τ > 0.3 mostly appear in July during 2006; in 2007 they are more widely distributed over time. Some PSCs are still detected during September, when the overall frequency of PSCs occurrence has already severely dropped.

Ancillary GMAO (GEOS-4) temperatures show PSCs with τ > 0.3 are correlated with minimum daily temperatures below -85°C south of 60°S except for a few outliers in September. Evaluating actual PSC temperatures is uncertain since GMAO temperatures may be too warm if gravity waves are involved (Sect. 3.2); extracting values at the coordinates and altitudes of PSCs gives an interval between -95°C and -80°C with a peak at -88°C, i.e. the cold end of the general PSC range (*Noel et al.* 2008).

Regarding microphysical properties, the detected PSCs strongly depolarize ($0.35 < \delta < 0.5$), with a very narrow distribution of depolarization ratios (Fig. 8), in stark contrast with the general PSC population, which exhibits a larger distribution of depolarization ratios centered on zero (Fig. 4 in *Noel et al.* 2008). Moreover, lidar ratios are in the $20 < S < 50$ range (not shown). Those results strongly suggest that these PSCs are in majority ice-based (Type II), consistent with their colder temperatures. These findings are stable in 2006 and 2007.

**5. Discussion**

The case study (Sect. 3) exhibits a strong correlation with very cold temperatures appearing during small-scale fluctuations, shown by mesoscale model to originate from orographic waves propagating from the Peninsula to the low and mid-stratosphere. This is consistent with previous studies that show orographic waves are not uncommon over this area between June and September (using e.g. MLS observations, *Jiang and Wu* 2002). Mountain waves have already been associated with the formation of PSCs in the Northern Hemisphere (e.g., *Carslaw et al.* 1998, *Dörnbrack et al.* 1999, *Voigt et al.* 2000). This kind of observations are more seldom in the Southern Hemisphere, where observing sites are fewer. Presented temperatures, either in ECMWF high-resolution reanalyses or using the WRF model (down to -100°C,173K), are cold enough to trigger ice formation at stratospheric levels of relative humidity (*Savigny et al.* 2005). The high depolarization and lidar ratios S ~ 20 of the case study PSC, coupled to cold temperatures, strongly suggest it is made of ice crystals (Type II). Moreover, a Type Ia PSC is observed downstream from this Type II PSC. *Carslaw et al.* (1999) described a mechanism where NAT particles can be rapidly formed by heterogeneous nucleation on ice particles of Type II PSCs, and later advected; this has since been observed numerous times over the North Pole (*Hu et al. 2002, Dornbrack et al.* 2002*, Luo et al.* 2003*, Voigt et al.* 2003*, Svendsen et al.* 2005*, Blum et al.* 2006). Over Antarctica, these mechanisms were also observed (*Cariolle et* *al.* 1989), but supposed to be secondary since the intravortex temperatures are cold enough to trigger NAT formation by themselves (*Eckermann et al.* 2006). However, MIPAS observations suggested that mountain-wave PSCs above the Peninsula, following the same mechanism, may have a long-range influence on the formation of NAT-based PSCs on synoptic scales inside the polar vortex (*Höpfner et al.* 2006b). We suggest the Type Ia PSC observed here could be generated from this same mechanism, since (1) it was observed downstream from the gravity-wave-generated Type II PSC, (2) back-trajectories show the air parcels that compose it went through temperatures colder than the ice frost point, and (3) no PSC is observed upstream where temperatures are similar. Since this configuration lasts two days, and vortex winds were intense (up to 80 m.s$^{-1}$ in ECMWF reanalyses and WRF3 results at ~10 hPa), its impact on stratospheric chemistry could be significant. A detailed microphysical modeling of the nucleation processes involved in the formation of this Type Ia PSC is however required to verify this hypothesis.

Comparisons with various temperature fields suggest that, in the context of PSC studies using CALIOP observations, temperatures from mesoscale simulations with high ceilings should be used (Sect. 3.2). Due to CALIOP's high spatial resolution, coarser temperature fields (e.g. GMAO) could overestimate temperatures by as much as 10 K depending on wind speed and gravity wave intensity, at least for PSCs similar to the case study (this applies in part to results from *Noel et al.* 2008). These findings might also be partly relevant to studies at other latitudes where small-scale dynamics interact with clouds (e.g. convection in the Tropics).

In austral winters 2007-2008, PSCs with optical depths above 0.3 are extremely rare (Sect. 4) and localized in the Western hemisphere, mostly over the Antarctic Peninsula. They are correlated with coldest minimum temperatures in GMAO fields, are overall colder than average PSCs, and very consistently produce high depolarization ratios, all of which strongly suggest they are ice-based (e.g. *Massoli et al.* 2006). It is not clear yet if all these PSCs are linked to gravity waves, but since their geographic concentration and optical properties (depolarization and lidar ratios) are very similar to the case study, the conditions which initiated the formation of the June 27$^{th}$, 2006 PSC could also apply. *Blum et al.* (2005) found using ground-based

14                                                                                                                                          14

lidar observations that, over Esrange (Sweden), ice PSCs were only observed coupled to gravity waves; but as temperatures are generally colder over the South Pole their conclusion might not apply here. Mesoscale modeling suggest that, in presence of orographic waves, global-scale reanalyses can overestimate temperatures in studied PSCs by 10 K (Sect. 3.2); an equivalent correction to the PSC temperature distribution (-95°C to -80°C or 178K to 193K) would bring their temperatures close to the case study (-105° to -90°C or 168K to 183K), at temperatures consistent with stratospheric ice formation (*Bacmeister et al.* 1990). Since this shows that a correlation of CALIOP's PSC observations with temperature requires the use of appropriately-tuned mesoscale models, future work involves running high-resolution simulations over Antarctica for one or more seasons, in order to evaluate reliably if the results of the case study can be generalized to other PSCs. Moreover, only two austral winters were analyzed here; it is unclear how representative they were in terms of mountain wave activity over Antarctica.

Regarding case studies, future work involves investigating the life cycle of particles after cloud formation. A certain amount of sedimentation occurs at PSC level (thus dehydration), but as the gravity wave is also associated with strong vertical winds it will also inject humidity from lower altitudes (which could have helped create the PSC in the first place). How gravity wave PSCs affect stratospheric humidity is therefore unclear. Moreover, the impact of PSCs similar to the present case study on stratospheric chemistry and other PSCs needs to be assessed, regarding e.g. their potential as initiator of Type I PSC formation. Applying a microphysical model of PSC particle growth on air mass trajectories going through temperatures from mesoscale simulations could shed some light on this issue. Such results should be interpreted in a broader context once the generality of the case study PSC is better known.

*Acknowledgments: We would like to thank Myrto Valari, Lionel Guez, Riwal Plougonven and Sophie Bastin for their help with the WRF model. The CALIOP level 1 and 2 data sets were obtained from the Climserv computing facility, the ICARE thematic center, and the NASA Langley Research Center Atmospheric Science Data Center.*

**Figures**

Fig. 1. June 27th, 2006 PSC case study. a) attenuated total backscatter observed from CALIOP; b) depolarization ratio after clear sky removal (relevant CALIPSO trajectory as an inset); c) lidar ratio S and d) optical depth (532 nm).

Fig. 2. Temperature fields extracted at the coordinates of CALIOP profiles along the case study orbit as a function of latitude and pressure, using a) GMAO (GEOS-4) reanalyses; b) GMAO (GEOS-5) reanalyses; c) NCEP reanalyses (2.5° resolution); d) ECMWF reanalyses (0.5° resolution) and results from the e) WRF1 f) WRF2 g) WRF3 simulations. Outlines of clouds with backscatter higher than $1.5 \ 10^{-3}$ km$^{-1}$.sr$^{-1}$ are superposed for visual reference.

Fig. 3. Temperature distributions for the case study (June 27th 2006), inside the PSC outline in Fig. 2, using values from a) GMAO (GEOS-4 and GEOS-5) reanalyses; b) NCEP (2.5° resolution) and ECMWF (0.5° resolution) reanalyses and results from the c) WRF1 d) WRF2 and e) WRF3 simulations.

Fig. 4. Atmospheric temperature (upper row), and zonal velocity (lower row) from the WRF3 simulation on June 27, 0300 UT: (left column) map at 40 hPa (~ 20 km), (middle column) cross section at 72°S, and (right column) disturbances at 72°S obtained by removing a second-order polynomial fit to the simulated fields at each latitude in the WRF domain. The 185 K iso-contour is displayed to roughly delimitate temperatures in WRF3 simulations that sustain ice formation (Assuming 5 ppmv of water vapor in the polar lower stratosphere, the frost point temperature is 189.4 K at 60 hPa, and 183.2 K at 20 hPa, *Marti and Mauersberger*, 1993). The track of CALIPSO orbit displayed in Figure 1 is shown with the bold dashed line on the maps (left column).

Fig. 5. Top panel: Atmospheric temperature disturbances at 20 hPa (left) and 60 hPa (right) from the WRF3 simulation on June 26, 1200 UT, when back-trajectories cross wave-induced low-temperature areas in the lee of the Antarctic Peninsula. The thick contours roughly indicate temperatures below $T_{ICE}$ (see text). The bold dashed lines show the track of CALIPSO orbits on June 27. Bottom panel: Attenuated Total Backscatter (left) and depolarization ratios (right) observed at 532 nm from CALIPSO

orbits 05-07-12 (West/upstream, top panel) and 01-49-27 (East/downstream, bottom panel), with backtrajectories intersections indicated with numbers. The relevant orbits and backtrajectories are shown on the top maps.

Fig. 6. Atmospheric conditions over the case study area on June 29th, 2006. Top: WRF3 atmospheric temperature map at 20 hPa with the two CALIPSO orbit trajectories shown below. Middle: attenuated total backscatter at 532 nm (left) observed from CALIOP closest to the case study on orbit 03-15-58, with the June 27th study PSC outlined in white, and vertical section of WRF3 temperature along the same orbit (right) with this June 29th PSC outlined in blue. Bottom: same as middle, for eastward, downstream orbit 01-37-03.

Fig. 7. Detections of PSCs with $\tau > 0.3$ in CALIOP profiles over Antarctica during 2006 and 2007 austral winters.

Fig. 8. Depolarization ratio distribution for PSCs with $\tau > 0.3$ during the 2006 and 2007 austral winters.

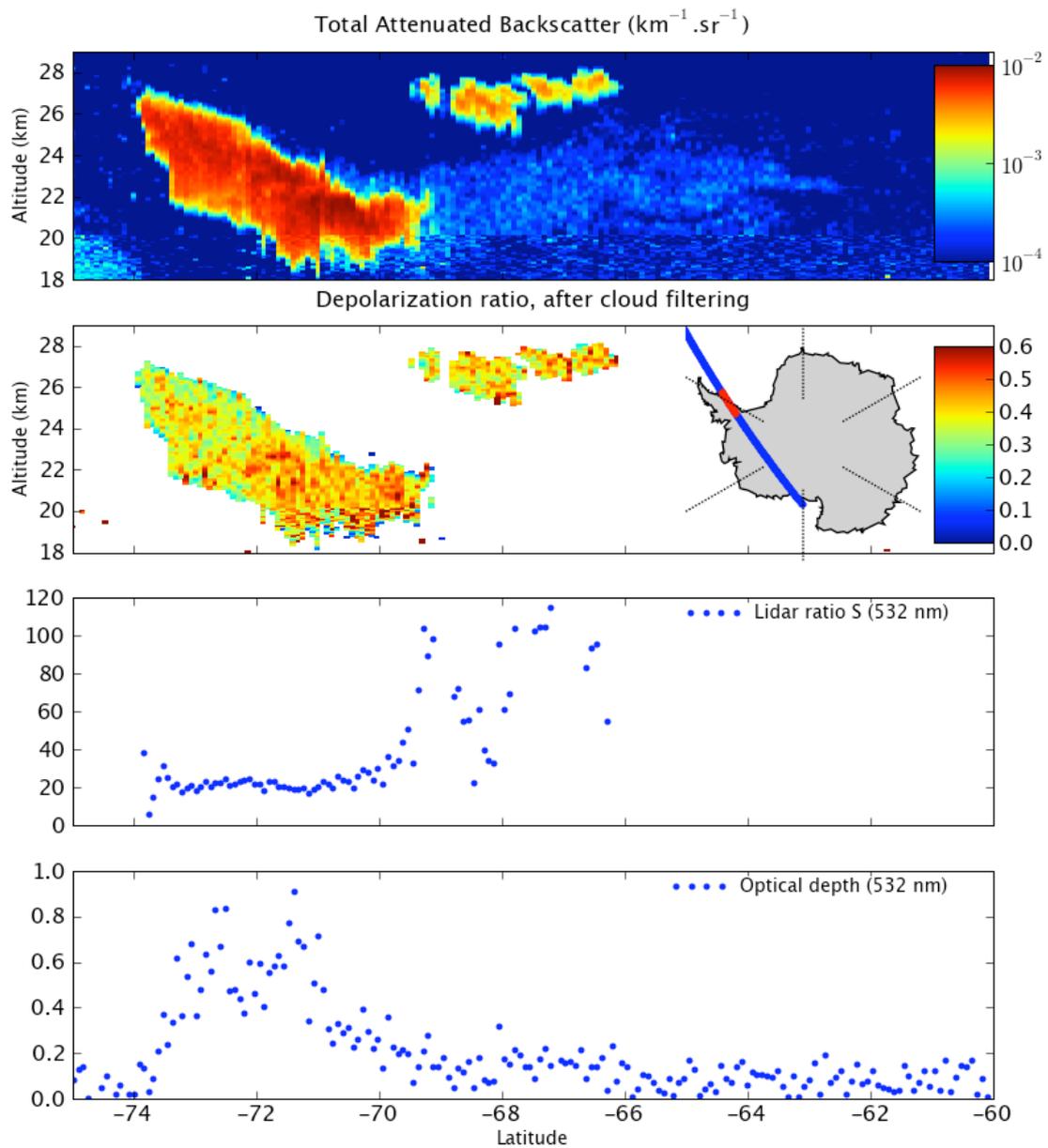

Fig. 1. June 27th, 2006 PSC case study. a) attenuated total backscatter observed from CALIOP; b) depolarization ratio after clear sky removal (relevant CALIOP trajectory as an inset); c) lidar ratio S and d) optical depth (532 nm).

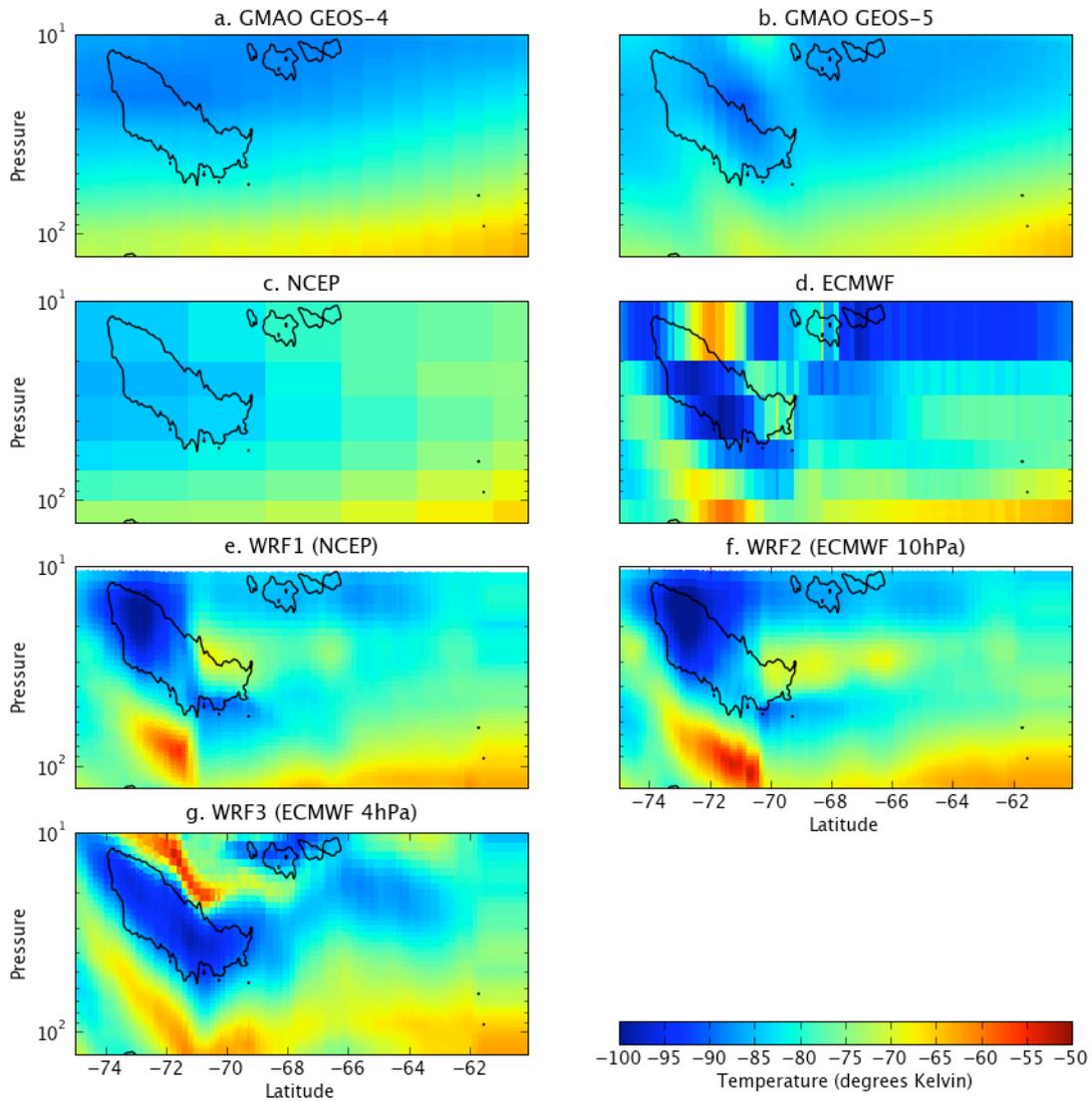

Fig. 2. Temperature fields extracted at the coordinates of CALIOP profiles along the case study orbit as a function of latitude and pressure, using a) GMAO (GEOS-4) reanalyses; b) GMAO (GEOS-5) reanalyses; c) NCEP reanalyses (2.5° resolution); d) ECMWF analyses (0.5° resolution) and results from the e) WRF1 f) WRF2 g) WRF3 simulations. Outlines of clouds with backscatter higher than $1.5 \cdot 10^{-3}$ km$^{-1}$.sr$^{-1}$ are superposed for visual reference. GMAO fields are coincident in time with CALIOP observations (June 27th, 2006 3:30 UTC), NCEP and ECMWF fields were extracted at midnight (closest reanalyses available), WRF results were extracted at 3:00 UTC (simulation time).

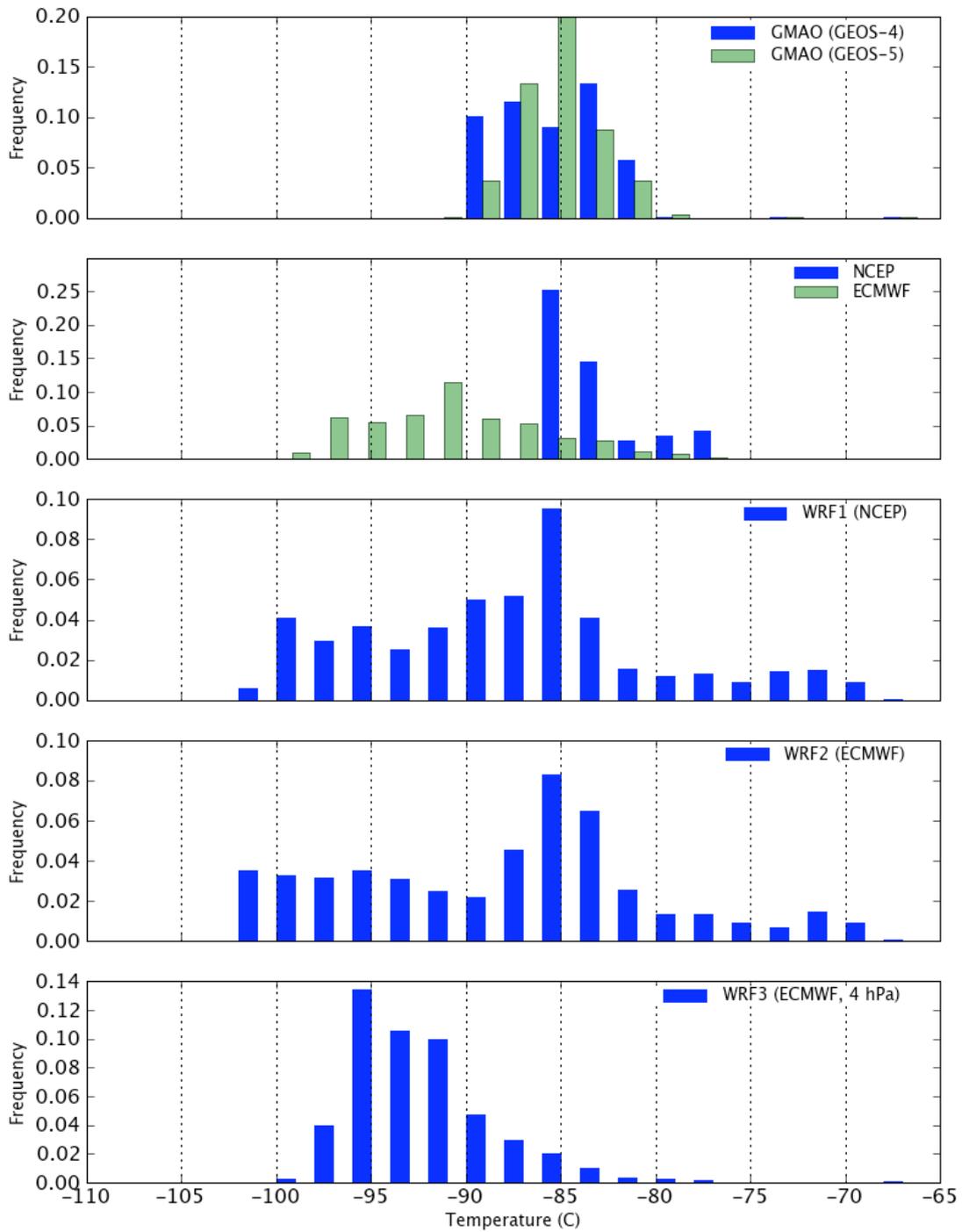

Fig. 3. Temperature distributions for the case study (June 27th 2006), inside the PSC outline in Fig. 2, using a) GMAO (GEOS-4 and GEOS-5) reanalyses; b) NCEP (2.5° resolution) and ECMWF (0.5° resolution) reanalyses and results from the c) WRF1 d) WRF2 and e) WRF3 simulations.

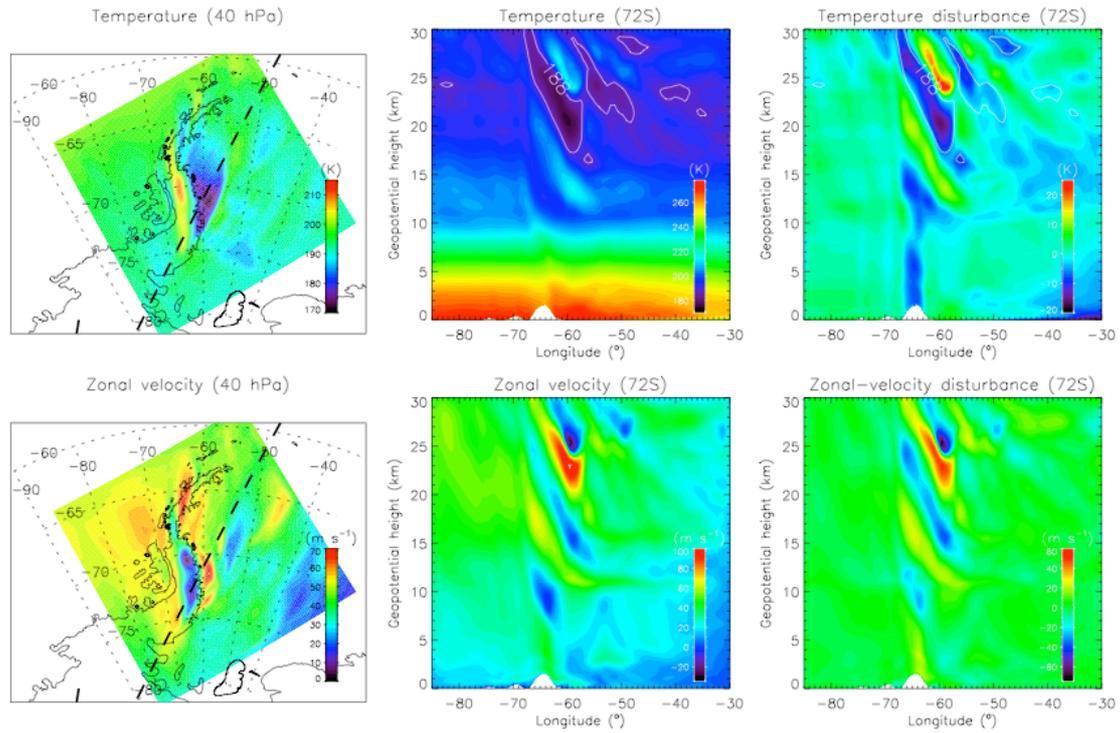

Fig. 4. Atmospheric temperature (upper row), and zonal velocity (lower row) from the WRF3 simulation on June 27, 0300 UT: (left column) map at 40 hPa (~ 20 km), (middle column) cross section at 72°S, and (right column) disturbances at 72°S obtained by removing a second-order polynomial fit to the simulated fields at each latitude in the WRF domain. The 185 K iso-contour is displayed to roughly delimitate temperatures in WRF3 simulations that sustain ice formation (Assuming 5 ppmv of water vapor in the polar lower stratosphere, the frost point temperature is 189.4 K at 60 hPa, and 183.2 K at 20 hPa, *Marti and Mauersberger*, 1993). The track of CALIPSO orbit displayed in Figure 1 is shown with the bold dashed line on the maps (left column).

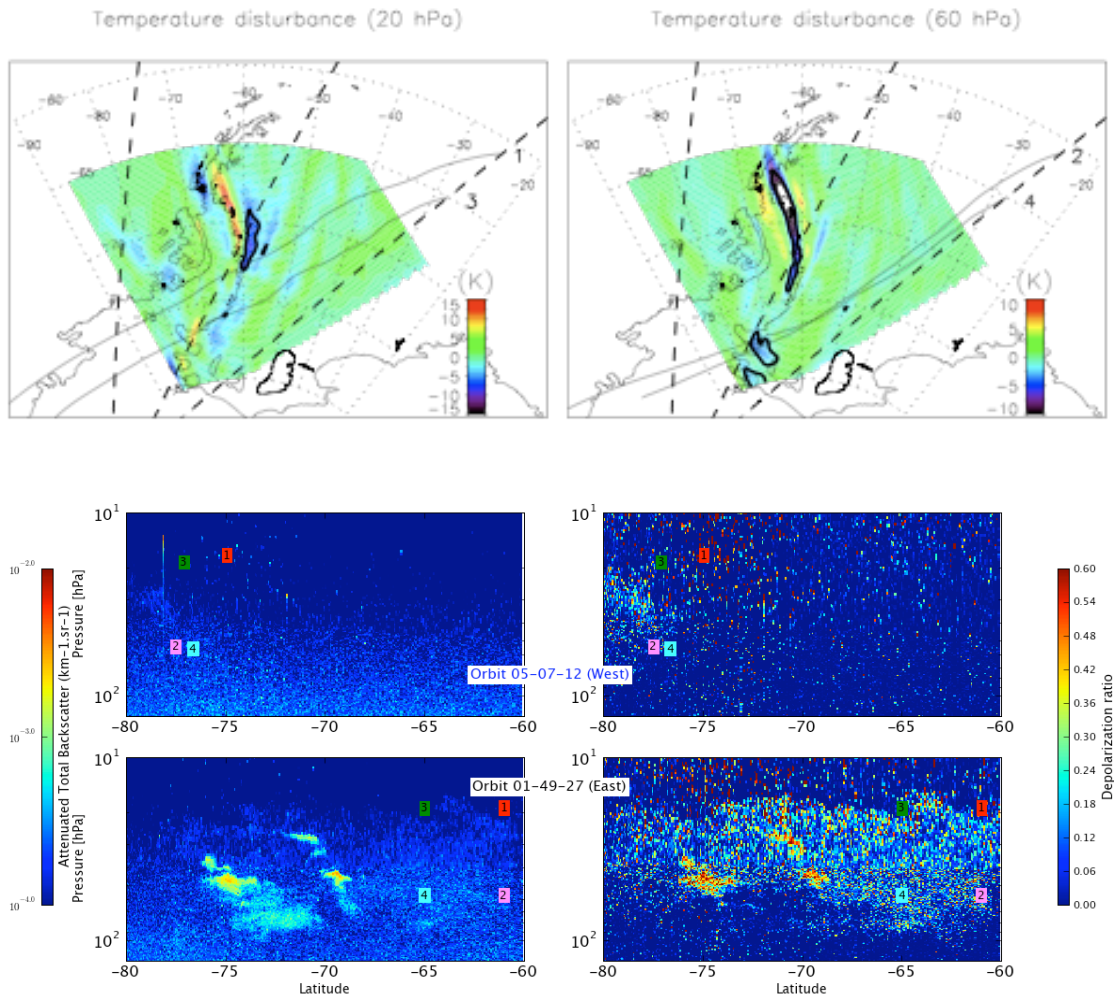

Fig. 5. Top panel: Atmospheric temperature disturbances at 20 hPa (left) and 60 hPa (right) from the WRF3 simulation on June 26, 1200 UT, when back-trajectories cross wave-induced low-temperature areas in the lee of the Antarctic Peninsula. The thick contours roughly indicate temperatures below $T_{ICE}$ (see text). The bold dashed lines show the track of CALIPSO orbits on June 27. Bottom panel: Attenuated Total Backscatter (left) and depolarization ratios (right) observed at 532 nm from CALIPSO orbits 05-07-12 (West/upstream, top panel) and 01-49-27 (East/downstream, bottom panel), with backtrajectories intersections indicated with numbers. The relevant orbits and backtrajectories are shown on the top maps.

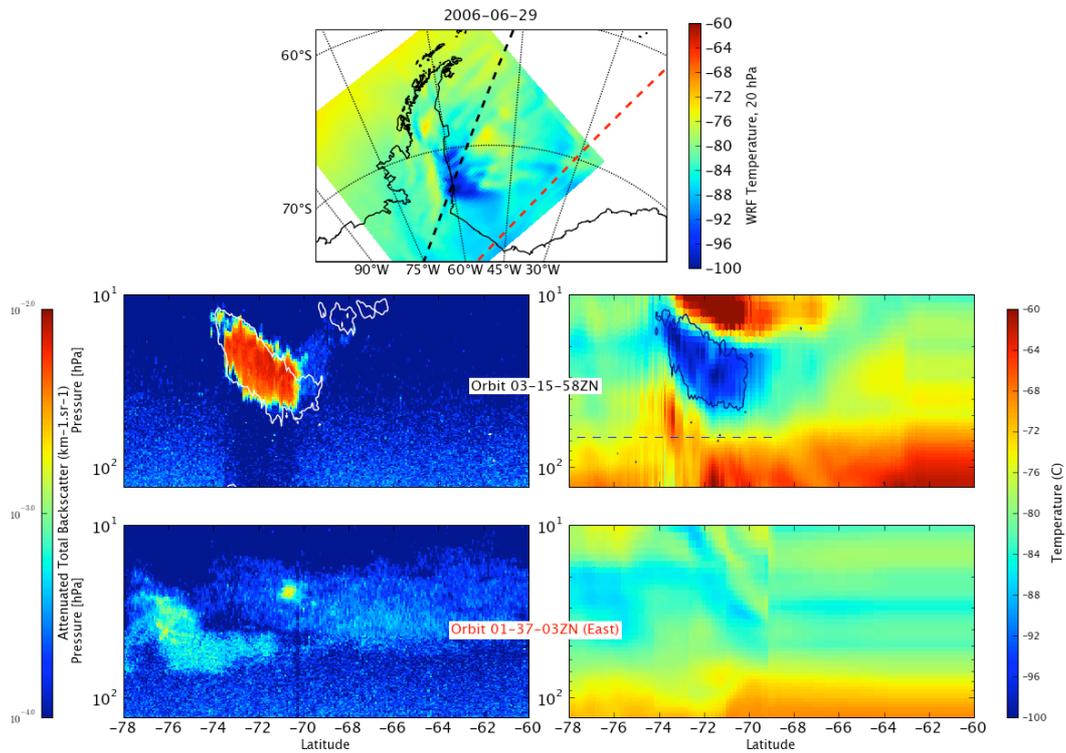

Fig. 6. Atmospheric conditions over the case study area on June 29th, 2006. Top: WRF3 atmospheric temperature map at 20 hPa with the two CALIPSO orbit trajectories shown below. Middle: attenuated total backscatter at 532 nm (left) observed from CALIOP closest to the case study on orbit 03-15-58, with the June 27th study PSC outlined in white, and vertical section of WRF3 temperature along the same orbit (right) with this June 29th PSC outlined in blue. Bottom: same as middle panel, for eastward, downstream orbit 01-37-03.

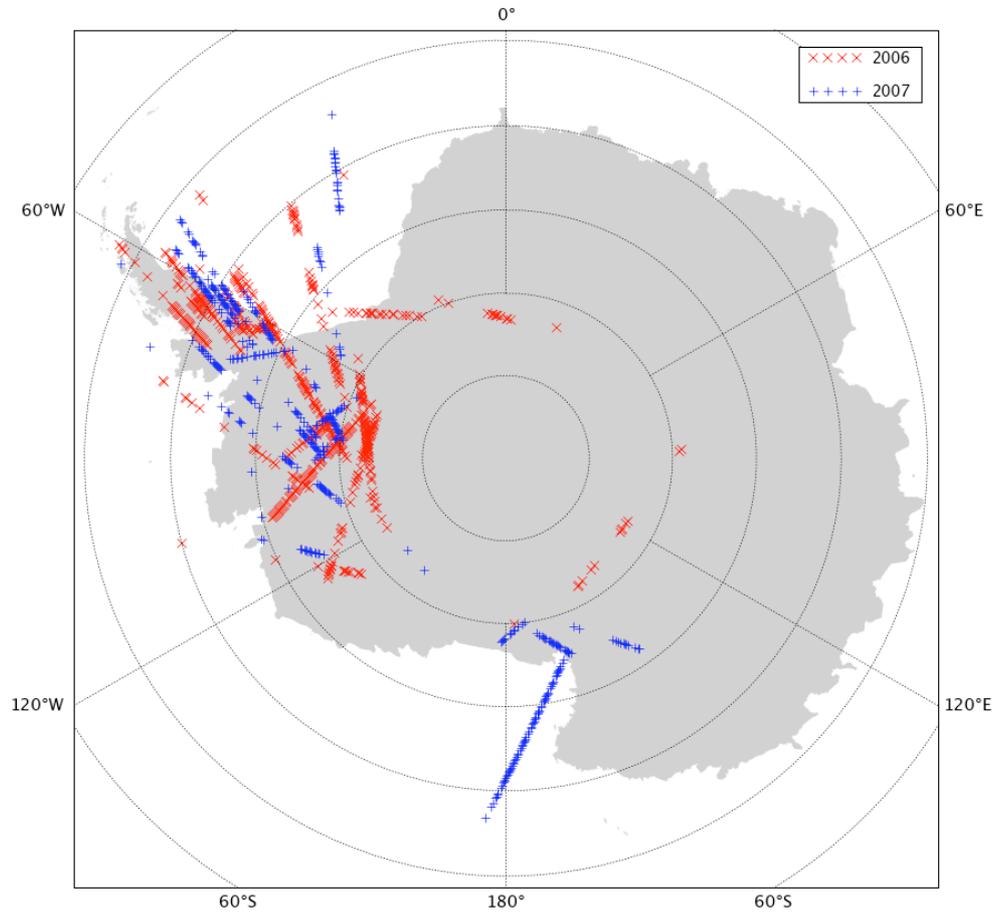

Fig. 7. Detections of PSCs with $\tau > 0.3$ in CALIOP profiles over Antarctica during 2006 and 2007 austral winters.

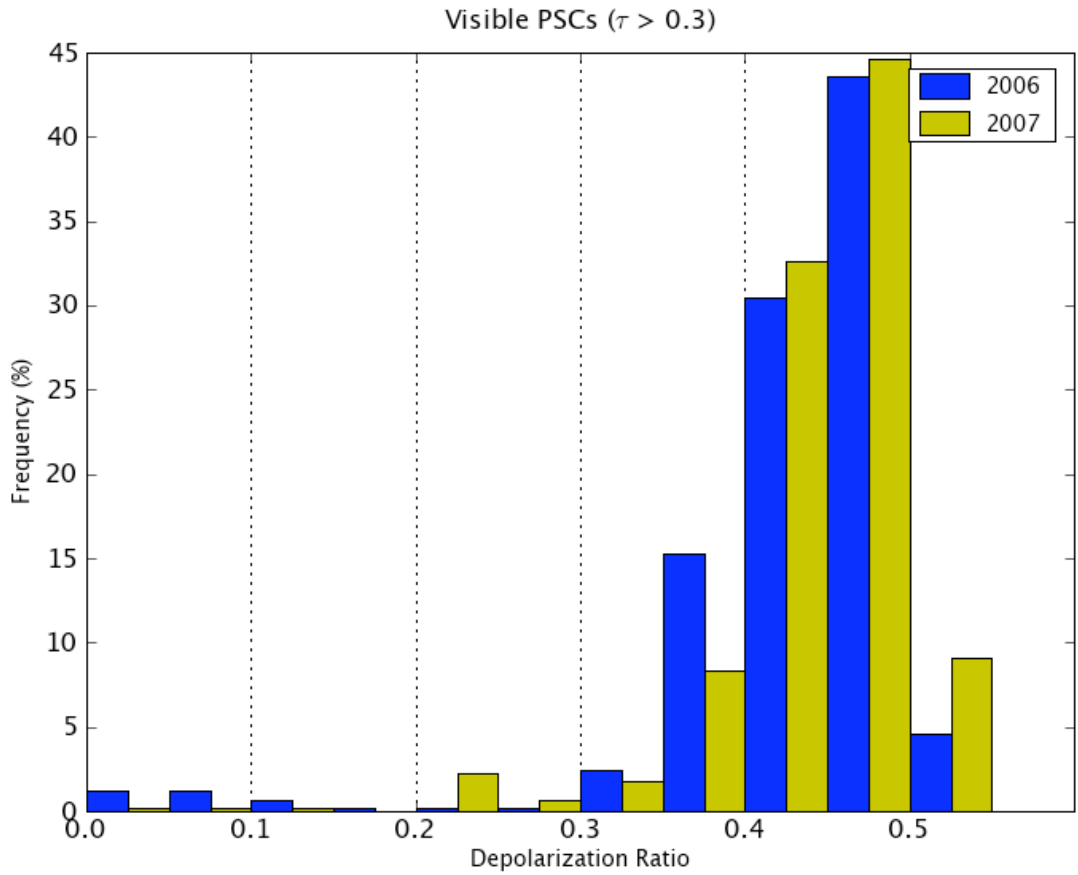

Fig. 8. Distributions of depolarization ratio for PSCs with τ > 0.3 during the 2006 and 2007 austral winters.